\documentclass[aps,prd,amssymb,amsmath,nobibnotes,nofootinbib,superscriptaddress, reprint, 12pt ]{revtex4}
\usepackage{amsfonts,amsmath,hyperref,url, color}
\allowdisplaybreaks
\usepackage{bm, bbm}
\usepackage{graphicx}
\usepackage{caption}
\usepackage{mathtools}
\usepackage{t1enc}
\usepackage{hepnames}
\usepackage{tensor}
\usepackage{comment}
\usepackage{float}

\usepackage{dcolumn}

\usepackage{adjustbox,booktabs}

\usepackage{makecell}
\usepackage{array}
\DeclareMathAlphabet\mathbfcal{OMS}{cmsy}{b}{n}

\usepackage{soul}

\newcommand{\bc}{\begin{center}}
\newcommand{\ec}{\end{center}}
\newcommand{\be}{\begin{eqnarray}}
\newcommand{\ee}{\end{eqnarray}}
\newcommand{\bs}{\begin{slide}}
\newcommand{\es}{\end{slide}}

\newcommand{\bi}{\begin{itemize}}
\newcommand{\ei}{\end{itemize}}

\begin{document}
\title{Finite $\kappa$-deformed two-particle boost}

\author{Andrea Bevilacqua}
\affiliation{National Centre for Nuclear Research, ul. Pasteura 7, 02-093 Warsaw, Poland}
\author{Jerzy Kowalski-Glikman}
\affiliation{University of Wroc\l{}aw, Faculty of Physics and Astronomy, pl.\ M.\ Borna 9, 50-204
Wroc\l{}aw, Poland}
\affiliation{National Centre for Nuclear Research, ul. Pasteura 7, 02-093 Warsaw, Poland}
\author{Wojciech Wi\'slicki}
\affiliation{National Centre for Nuclear Research, ul. Pasteura 7, 02-093 Warsaw, Poland}

\date{\today}

\begin{abstract}
In this paper we consider the $\kappa$-deformed boost acting on a two-particles states. Using techniques developed in the case of infinitesimal boost we compute explicit expression for the components of the finite boost matrices acting on the first and second particle. We briefly discuss phenomenological consequences of our findings.

\end{abstract}

\maketitle

\section{Introduction}

It is commonly believed that when  quantum gravity effects are taken into account the standard description of quantum particles and fields is dramatically modified \cite{Doplicher:1994tu}, \cite{Hossenfelder:2012jw}, leading to new effects that might be even observable with current or near future technologies \cite{AmelinoCamelia:2008qg}, \cite{Addazi:2021xuf}. One class of these effects is associated with possible deformations of spacetime symmetries (see \cite{Arzano:2021scz} for a recent comprehensive review).
Interesting effects originate also from deformation of discrete transformations: the charge conjugation C, space inversion P and time reversal T.
Deformation of these symmetries can be expected at high energy and result in measurable effects, e.g. on particle lifetimes \cite{arzano_kowalski_wislicki_2022}.
As known from research on the violation of these symmetries and their combinations in weak interactions, other observables that are very sensitive to their violation require measurements of multi-particle states.
In particular, subtle effects of the CP violation, or even more elusive CPT non-conservation anticipated in some extensions of the Standard Model, could be best observed in interference of two or more particles \cite{cpcpt_interference}. 
It is therefore important to build a  theoretical framework for description of deformed states of more than one particle boosted to very high energy.  

The presence of quantum deformations of spacetime symmetries was firmly established and is well understood in the context of gravity in 2+1 spacetime dimensions, see e.g., \cite{Matschull:1997du,Bais:2002ye,Freidel:2005me,Dupuis:2020ndx}. In physical 3+1 dimensions the situation is less clear, but  there are indications that also in this case the spacetime symmetries should get deformed, as a consequence of quantum gravitational effects \cite{Freidel:2003sp}. In what follows we will consider a particular, best studied example of a quantum deformation, called $\kappa$-Poincar\'e  \cite{Lukierski:1993wx}, \cite{Majid:1994cy}. 

One of the characteristic features of quantum deformations of spacetime symmetries is the emergence of a nontrivial co-product, i.e., the rules of action of these symmetries on multiparticle states. Take the momentum operator $\mathcal{P}$ as an example. The one-particle state $|P\rangle$ is its eigenstate 
\begin{align}
    \mathcal{P}\triangleright |P\rangle = P |P\rangle
\end{align}
Thus the momentum operator computes momenta of the particles. More abstractly, the momentum operator maps one-particle Hilbert space to itself $\mathcal{P}: \mathcal{H}\rightarrow\mathcal{H}$. This however does not provide any canonical way in which the translation operator acts on two-particle state, an element of the tensor product of two one-particle Hilbert spaces. This action is captured by the so called coproduct
\begin{align}
 \Delta\mathcal{P}: \mathcal{H}\otimes\mathcal{H}\rightarrow\mathcal{H} \otimes\mathcal{H}  
\end{align}
The simplest possibility, used in the standard undeformed quantum field theory is the Leibnizean rule
\begin{align}
   \Delta\mathcal{P}_{Leibniz} \triangleright |P\rangle\otimes|Q\rangle = \left(\mathcal{P} \triangleright |P\rangle\right)\otimes|Q\rangle + |P\rangle\otimes\left(\mathcal{P} \triangleright|Q\rangle\right) = (P+Q)|P\rangle\otimes|Q\rangle
\end{align}
Thus $\Delta\mathcal{P}$ computes the total momentum of the two-particle system. This can be extended to the multi-particles' case by associativity: 
$$
 \Delta\mathcal{P} \triangleright |P^{(1)}\rangle\otimes|P^{(2)}\rangle\otimes|P^{(3)}\rangle =  \Delta\mathcal{P} \triangleright \left(|P^{(1)}\rangle\otimes|P^{(2)}\rangle\right)\otimes|P^{(3)}\rangle
$$
In the case of $\kappa$-Poincar\'e deformation, for time and space components we find (in the classical basis, in which $P_\mu$ form a standard Lorentz vector) 
\begin{align}
   \Delta\mathcal{P}_{\kappa}{}_0 \triangleright |P\rangle\otimes|Q\rangle &= \left(\frac1\kappa\, P_0( Q_0 +Q_4) + \frac{\mathbf{P}\cdot \mathbf{Q}}{P_0+P_4} + Q_0\, \frac\kappa{P_0+P_4}\right)|P\rangle\otimes|Q\rangle \\
   \Delta\mathbf{\mathbfcal{P}}_{\kappa}{} \triangleright |P\rangle\otimes|Q\rangle &=\left(\frac1\kappa\, \mathbf{P}\, (Q_0+Q_4) + \mathbf{Q}\right) |P\rangle\otimes|Q\rangle 
\end{align}
where we use boldface to denote the spacial components of a four-vector and the fourth component is defined as $P_4=\sqrt{\kappa^2 +P_0^2 -\mathbf{P}^2}$. In the recent paper \cite{Arzano:2022ewc}  a new, group-theoretical method of deriving the action of deformed $\kappa$-Poincar\'e symmetries on two-particle states was introduced. In the case of Lorentz symmetry the formulas describing infinitesimal boosts was derived, and here we will extend this construction to the case of finite Lorentz boosts.

\section{Action of deformed boosts on two-particle system}\label{SectII}

Following \cite{Arzano:2022ewc}, to derive the action of Lorentz boosts  on two-particle state, we start with the Iwasawa decomposition of the algebra $\mathfrak{so}(1,4)$  
\begin{align}
	\mathfrak{so}(1,4) = \mathfrak{so}(1,3)\oplus \mathfrak{a} \oplus \mathfrak{n}
\end{align}
where $\mathfrak{a}$ and $\mathfrak{n}$ are generated respectively by
\begin{align}\label{x0xi}
	x^0
	=
	-\frac{i}{\kappa}
	\begin{pmatrix}
		0 & \bm{0} & 1 \\
		\bm{0} & \bm{0} & \bm{0} \\
		1 & \bm{0} & 0
	\end{pmatrix}
\qquad
	\mathbf{x}
	=
	\frac{i}{\kappa} 
	\begin{pmatrix}
		0 & \epsilon^T & 0 \\
		\epsilon & \bm{0} & \epsilon \\
		0 & -\epsilon^T & 0
	\end{pmatrix}.
\end{align}
and form the Lie algebra of the Lie group $AN(3)$ representing the deformed momenta (see \cite{Arzano:2021scz} for details). The  $\mathfrak{so}(1,3)$ represents Lorentz trasformations.

As the consequence of the Iwasawa decomposition, any $SO(4,1)$  element $G$ can be uniquely written as a group product $G =L g$ with $L\in SO(1,3)$ (Lorentz group) and $g\in AN(3)$ (momentum space group-manifold \cite{Arzano:2021scz}). Then there is a unique couple, $g'\in AN(3), L'_g\in SO(1,3)$ such that 
\begin{align}\label{iwasawa}
G=	L g = g' L'_g.
\end{align}
Notice that the Lorentz matrix $L_g$ on the right hand side can in general be momentum dependent. This matrix  will be of fundamental importance in discussion of the two-particle boost, and we need to find explicitly its components. To this end, we use the representation \eqref{x0xi} to write both $g$ and $g'$, and consider without loss of generality $L$ as a boost in the $x^1$ direction, to obtain
\begin{align}
	L
	=
	\begin{pmatrix}
		\gamma & -\beta\gamma & 0 & 0 & 0 \\
		-\beta\gamma & \gamma & 0 & 0 & 0 \\
		0 & 0 & 1 & 0 & 0 \\
		0 & 0 & 0 & 1 & 0 \\
		0 & 0 & 0 & 0 & 1
	\end{pmatrix}
\qquad
	g
	=
	\begin{pmatrix}
		\tilde{P}_4 & \kappa\mathbf{P}_1/P_+ & \kappa\mathbf{P}_2/P_+ & \kappa\mathbf{P}_3/P_+ & P_0 \\
		\mathbf{P}_1 & \kappa & 0 & 0 & \mathbf{P}_1 \\
		\mathbf{P}_2 & 0 & \kappa & 0 & \mathbf{P}_2 \\
		\mathbf{P}_3 & 0 & 0 & \kappa & \mathbf{P}_3 \\
		\tilde{P}_0 & -\kappa\mathbf{P}_1/P_+ & -\kappa\mathbf{P}_2/P_+ & -\kappa\mathbf{P}_3/P_+ & P_4
	\end{pmatrix}\label{gmatrix}
\end{align}
\begin{align}
	L'_g
	=
	\begin{pmatrix}
		L_{00} & L_{01} & L_{02} & L_{03} & 0 \\
		L_{10} & L_{11} & L_{12} & L_{13} & 0 \\
		L_{20} & L_{21} & L_{22} & L_{23} & 0 \\
		L_{30} & L_{31} & L_{32} & L_{33} & 0 \\
		0 & 0 & 0 & 0 & 1
	\end{pmatrix}
\qquad
	g'
	=
	\begin{pmatrix}
		\tilde{P}'_4 & \kappa\mathbf{P}'_1/P'_+ & \kappa\mathbf{P}'_2/P'_+ & \kappa\mathbf{P}'_3/P'_+ & P'_0 \\
		\mathbf{P}'_1 & \kappa & 0 & 0 & \mathbf{P}'_1 \\
		\mathbf{P}'_2 & 0 & \kappa & 0 & \mathbf{P}'_2 \\
		\mathbf{P}'_3 & 0 & 0 & \kappa & \mathbf{P}'_3 \\
		\tilde{P}'_0 & -\kappa\mathbf{P}'_1/P'_+ & -\kappa\mathbf{P}'_2/P'_+ & -\kappa\mathbf{P}'_3/P'_+ & P'_4
	\end{pmatrix}\label{gmatrix'}
\end{align}
where we abbreviate
\begin{align}\label{tilde}
	\tilde{P}_4 = P_4 + \frac{\mathbf{P}^2}{P_+}
	\qquad
	\tilde{P}_0 = P_0 - \frac{\mathbf{P}^2}{P_+}
    \qquad
    P_+=P_0+P_4.
\end{align}
It can be checked \cite{Arzano:2022ewc} that if the entries $P_0, \mathbf{P}$ in the last column of the $g$-matrix in \eqref{gmatrix} represents the momenta before boost, the entries $P'_0, \mathbf{P}'$ in the last column of the $g'$-matrix in \eqref{gmatrix'} represents the boosted momenta.
After straightforward but tedious computations using eq. \eqref{iwasawa}, one obtains an explicit expression for the entries of the matrix $L'_g$, which can be found in Appendix \ref{appA}. 

When discussing possible experimental signatures of $\kappa$-deformation it is usually sufficient to consider only the first order approximation of the above formulas, proportional to $1/\kappa$ . The expansion of $L'_g$ up to first order in $1/\kappa$ can be written as a the sum of the undeformed  $\kappa\rightarrow\infty$ matrix and the first order one
\begin{align}\label{Lexp}
	L'_g
	=
	L'_g(\kappa\rightarrow\infty)
	+
	\frac{1}{\kappa} \tilde{L}'_g + O\left(\frac1{\kappa^2}\right) 
\end{align}
where
\begin{align}\label{L}
	L'_g(\kappa\rightarrow\infty)
	=
	\begin{pmatrix}
		\gamma & -\beta\gamma &	0 & 0 & 0 \\
		-\beta\gamma & \gamma & 0 & 0 & 0 \\
		0 & 0 & 1 & 0 & 0 \\
		0 & 0 & 0 & 1 & 0 \\
		0 & 0 & 0 & 0 & 1
	\end{pmatrix}
\end{align}
\begin{align}\label{Lpg}
	\tilde{L}'_g
	=
	\begin{pmatrix}
		(1-\gamma)
		\left[
		(1+\gamma) P_0-\beta\gamma \mathbf{P}_1
		\right] 
		& 
		\gamma\mathbf{P}_1
		-
		\gamma(\gamma\mathbf{P}_1 - \beta\gamma P_0)
		&
		(\gamma-1) \mathbf{P}_2
		&
		(\gamma-1) \mathbf{P}_3
		&
		0 \\
		\gamma\mathbf{P}_1
		-
		\gamma(\gamma\mathbf{P}_1 - \beta\gamma P_0)
		&
		\beta\gamma(\gamma-1) \mathbf{P}_1
		-
		\beta^2\gamma^2P_0 
		&
		-\beta \gamma \mathbf{P}_2
		&
		-\beta \gamma \mathbf{P}_3
		&
		0 \\
		(1-\gamma) \mathbf{P}_2
		&
		\beta\gamma \mathbf{P}_2
		&
		0
		&
		0 & 0 \\
		(1-\gamma) \mathbf{P}_3
		&
		\beta\gamma \mathbf{P}_3
		&
		0 & 0 & 0 \\
		0 & 0 & 0 & 0 & 0
	\end{pmatrix}
\end{align}
Notice that if $L'_g$ is a legitimate Lorentz transformation, then by definition one must have
\begin{align}
	L'_g \eta {L'}^T_g = \eta
\end{align}
where $\eta $ is the Minkowski metric. We know that the zeroth order in perturbation satisfies the relation
\begin{align}
	L'_g(\kappa\rightarrow\infty) \eta {L'}^T_g(\kappa\rightarrow\infty) = \eta
\end{align}
by definition, which means that we must have (ignoring terms of order $1/\kappa^2$)
\begin{align}
	L'_g(\kappa\rightarrow\infty) \eta (\tilde{L}')^T_g + \tilde{L}'_g \eta [L'_g(\kappa\rightarrow\infty)]^T = 0.
\end{align}
This is already ensured by the Iwasawa decomposition, but one can nevertheless explicitly check that \eqref{Lexp} satisfies this relations.

Let us now turning to computing the action of boost on a two-particle state \cite{Arzano:2022ewc}, \cite{Arzano:2022vmh}. By the same reasoning as before, we have 
\begin{align}
	(gh)' = L g  (L'_g)^{-1} L'_g h ({L'}_{hg})^{-1}
\end{align}
Notice that multiplication of the matrices $L g, L'_g h$ from the right by an arbitrary Lorentz matrix does not change the entries of their last columns so that $|(L g  (L'_g)^{-1})(P)\rangle = |(L g  )(P)\rangle$ and in writing the states we can neglect the Lorentz group elemet on the right of $g$. Therefore we define the action of the Lorentz transformation $\mathcal{L}$ on a two particle state  as 
\begin{align}\label{coprodboost}
\Delta\mathcal{L} \triangleright |g(P) \rangle\otimes |h(Q) \rangle=	|(L g)(P) \rangle |(L'_g h)(Q) \rangle.
\end{align}
where $(L g)(P)$ denotes (in the case of the boost along the first axis) the four top components of the last column of the product of the matrices $L$ and $g$ in \eqref{gmatrix}, in components explicitly  $(\gamma P_0 -\beta \gamma P_1, -\beta\gamma P_0 + \gamma P_1, P_2, P_3)$. The components of $L'_g h$ can be similarly computed using \eqref{Lexp} in the linearized case or using the formulas presented in the Appendix \ref{appA} in the full case.

\section{Two-particles kinematics}

Consider decay of a particle of mass $M$, originally at rest, into two particles of mass $m$. Starting from the first order expansion in $1/\kappa$ of the four-momentum deformed composition rule $(P\oplus Q)_0$ and $(P\oplus Q)_i$ we get 
\begin{align}
	(P\oplus Q)_0
	\approx
	P_0 + Q_0 + \frac{\mathbf{P}\mathbf{Q}}{\kappa} 
 \qquad
	(P\oplus Q)_i
	\approx
	P_i + Q_i + \frac{P_i Q_0}{\kappa}
\end{align}
and we need to impose 
\begin{align}\label{ident}
	P_0 + Q_0 + \frac{\mathbf{P}\mathbf{Q}}{\kappa}
	=
	M
\qquad
	P_i + Q_i + \frac{P_i Q_0}{\kappa}
	=
	0.
\end{align}
We need to find the spatial momenta such that these two relations are satisfied.
The conservation of spatial momenta tells us that the momenta $\mathbf{P}$ and $\mathbf{Q}$ are parallel, which means that we can align them with one of our axis, and we can write $P_\mu = (P_0,P,0,0)$ and $Q_\mu=(Q_0,Q,0,0)$ without loss of generality. We have therefore
\begin{align}
	P_0 + Q_0 + \frac{PQ}{\kappa}
	=
	M
\qquad
	P + Q + \frac{P Q_0}{\kappa}
	=
	0.
\end{align}
From the conservation of momenta, one can get the relation
\begin{align}\label{pq}
	P \approx
	-Q + \frac{Q Q_0}{\kappa}
\end{align}
which substituted back into the sum of energies gives (at first order in $1/\kappa$)
\begin{align}\label{Q}
	Q
	\approx
	\sqrt{M^2-4m^2}
	\left(
	\frac{1}{2}
	+
	\frac{3M}{8\kappa}
	\right)
\end{align}
Substituting back into \eqref{pq}, we also get
\begin{align}\label{P}
	P
	\approx
	-\sqrt{M^2-4m^2}
	\left(	
	\frac{1}{2}
	+
	\frac{M}{8\kappa}
	\right).
\end{align}
Notice that the modulus $P$ differs from that of $Q$ by $\sqrt{M^2-4m^2}\, M/4\kappa$.

We can now boost the two particles. Since the boost is in the $x$ direction but the two particles can go in arbitrary direction in the center of mass frame, we consider spherical coordinates with the polar axis oriented along the $x$ axis, so that the coordinates of the momenta $P_\mu$ and $Q_\mu$ are given by
\begin{align}
	P_\mu &= (P_0, P\cos\theta, P\sin\theta \sin\phi, P\sin\theta \cos\phi) =: (P_0, P_1, P_2, P_3) \\
	Q_\mu &= (Q_0, Q\cos\theta, Q\sin\theta \sin\phi, Q\sin\theta \cos\phi) =: (Q_0, Q_1, Q_2, Q_3).
\end{align}
When computing the boost, we have two possibilities, depending on which particle is the first and which is second in \eqref{coprodboost}, namely $L\triangleright P_\mu$ and $L'_g\triangleright Q_\mu$, or $L\triangleright Q_\mu$ and $L'_g\triangleright P_\mu$. Since the undeformed boost acts on momenta as usual, we will only concentrate in the action of $L'_g$. Furthermore, since the second case can be obtained from the first one by simply switching $P\longleftrightarrow Q$, we will only consider the first one. Using equations \eqref{L}, \eqref{Lpg}, \eqref{Q}, \eqref{P}, we can write down the first order expansion in powers of $1/\kappa$ of the momenta boosted using $L'_g$. We have
    \begin{align}
    \label{Qcoordinates}
		(L'_g\triangleright Q)_0
		&=
		\gamma(Q_0-\beta  {Q}_1)
        +
        \frac{1}{\kappa}
        \Big[
        (\gamma-1)( {P}_2 {Q}_2 +  {P}_3 {Q}_3 - \gamma  {P}_1 {Q}_1 + \beta\gamma Q_0  {P}_1) \nonumber \\
        &+
        P_0(Q_0 - Q_0\gamma^2 + \beta\gamma^2  {Q}_1)
        \Big]
		\\ 
		(L'_g\triangleright Q)_1
		&=
        \gamma( {Q}_1-\beta Q_0)
        +
        \frac{1}{\kappa}
        \Big[
        -\beta\gamma( {P}_2 {Q}_2) +  {P}_3 {Q}_3 -  {P}_1(Q_0 - \beta {Q}_1)(\gamma-1)\gamma \nonumber \\
        &+
        P_0( {Q}_1- {Q}_1\gamma^2 + Q_0\beta\gamma^2)
        \Big]
		\\ 
		(L'_g\triangleright Q)_2
		&=
         {Q}_2 
        +
        \frac{1}{\kappa}
        [ {P}_2(Q_0-Q_0\gamma +  {Q_1}\beta\gamma)]
		\\ 
  \label{Qcoordinates1}
		(L'_g\triangleright Q)_3
		&=
         {Q}_3 
        +
        \frac{1}{\kappa}
        [ {P}_3(Q_0-Q_0\gamma +  {Q_1}\beta\gamma)]
	\end{align}
One can also explicitly write the expressions of $L\triangleright P$ and $L'_g\triangleright Q$ obtained by substituting eq. \eqref{Q}, \eqref{P} in the formulae, these expressions can be found in Appendix \ref{appB}. Furthermore, we can also plot the angular dependence of the modulus of the deformed boosted momenta to compare it to the undeformed case. In figures below all the quantities on the axes are expressed in GeV. 

\begin{figure}[H]
	\centering
	\includegraphics[width=0.6\textwidth]{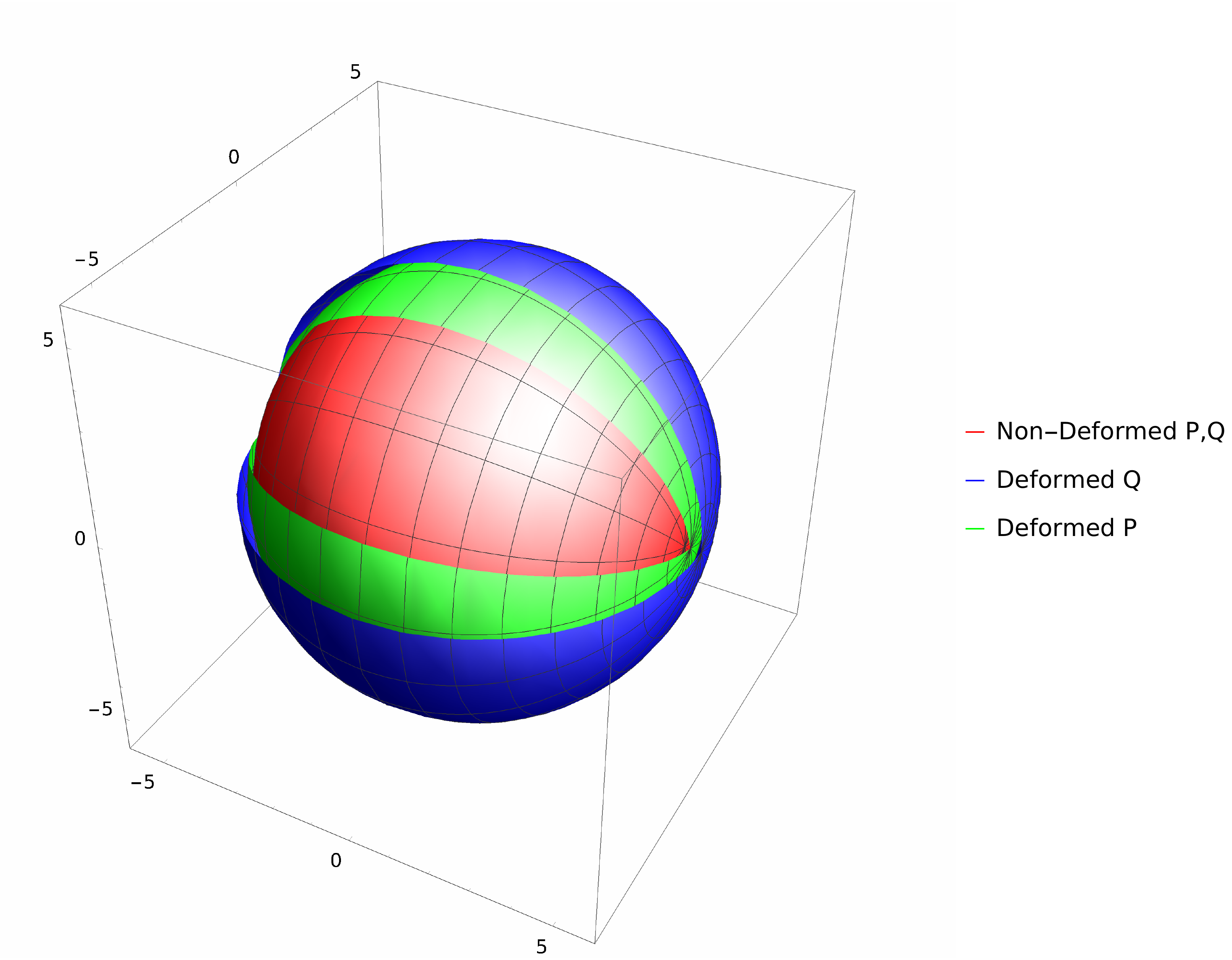}
    \caption{Angular distribution of deformed and undeformed momenta in the COM frame, with parameters $M=10 \;\mbox{GeV}$, $m=0.1 \;\mbox{GeV}$, $\kappa=10^2 \;\mbox{GeV}$. Notice the onion-like structure, with the undeformed momentum being innermost, followed by the deformed $P$, and finally the deformed $q$. Here, $P$ and $Q$ are antipodal (see eq. \eqref{ident}). To give a more concrete numerical estimate, the modulus of the undeformed momentum is $5.00$ GeV, the modulus of $P$ is $5.12$ GeV, while the modulus of $Q$ is $5.37$ GeV.}
    \label{fig:PQcomFig}
\end{figure}

In the center of mass (COM) frame, the two momenta are distributed on a sphere. Because of deformation, however, the spatial momenta $P$ and $Q$ are different in modulus (see eq. \eqref{Q}, \eqref{P}). In Figure \ref{fig:PQcomFig} we show the distribution of momenta in the COM frame, together with the distribution in the undeformed case. To highlight the qualitative difference, for the next plots we chose hypothetically $M=10 \;\mbox{GeV}$, $m=0.1 \;\mbox{GeV}$, $\kappa=10^2 \;\mbox{GeV}$. Furthermore, we chose the domains $\phi\in \left[-\frac{\pi}{10}, \frac{3\pi}{2}+\frac{\pi}{10}\right]$ and $\phi\in \left[0, \frac{3\pi}{2}\right]$ for the distributions of the deformed $P$ and the deformed $Q$ respectively, in order to clearly show how they are related to the undeformed one. Notice that the undeformed case contains only a single surface because both spatial momenta have the same modulus. 

\begin{figure}[H]
	\centering
	\includegraphics[width=0.88\textwidth]{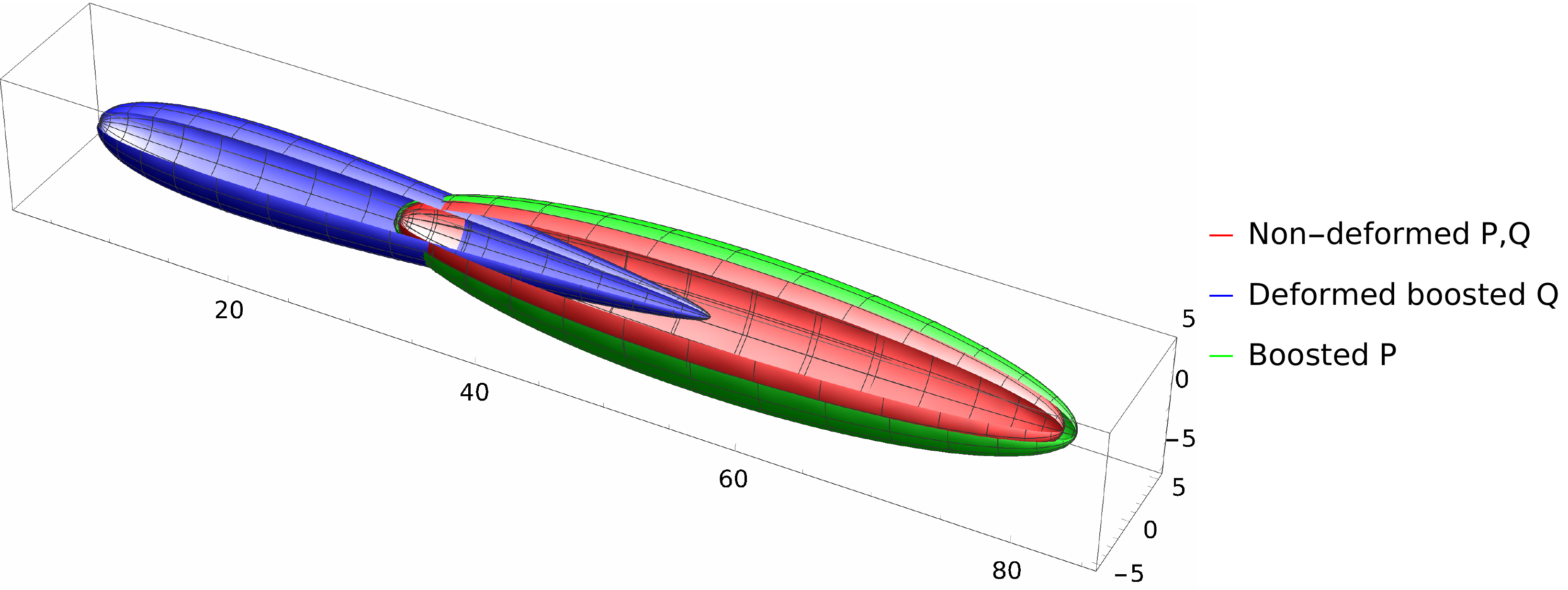}
    \caption{Angular distribution of deformed and undeformed momenta in the boosted frame, with parameters $\gamma=5$, $M=10 \;\mbox{GeV}$, $m=0.1 \;\mbox{GeV}$, $\kappa=10^2 \;\mbox{GeV}$. Recall that both $P$ and $Q$ feel the effects of deformation already in the modulus (see eq. \eqref{P} and \eqref{Q}). When boosting, $P$ gets boosted with a canonical non-deformed boost, while $Q$ is instead boosted in a deformed way.}
    \label{fig:PQboostFig}
\end{figure}

We then boost our momenta. In particular, $P_\mu$ will be canonically boosted while $Q_\mu$ will be boosted in a deformed way (recall however that both $P_\mu$ and $Q_\mu$ still have some effects of deformation in their modulus). In this case we chose $\gamma=5$ while the other parameters remain the same. We also restricted the domain in $\phi$ of all distributions to better show the features of the surfaces, and we obtain Figure \ref{fig:PQboostFig}. Notice that the deformed boost not only modifies the amplitude of the boosted momentum, but also its angular distribution.

\section{Conclusions}

In this paper we discussed the action of $\kappa$-deformed boost on a two-particles state. As a result of the deformation of boost co-product there is a $\kappa$-dependent difference between the way the first and second particle is boosted, depicted in Figure \ref{fig:PQboostFig}.

Our findings can have important implications for future research on deformed fields using two-particle correlations. 
It is common to study interference patterns in variables defined as a difference of particles' momenta or difference of their decay times.
Sensitivity of these variables to new phenomena is the best for their small values where interference pattern variates largely and statistics of data samples is sizable.
As can be seen from Eqs. (\ref{Qcoordinates}-\ref{Qcoordinates1}), for $P$ and $Q$ close to each other, their difference mostly depends on the $1/\kappa$ correction term.
This fact can be exploited in the phenomenological studies of interference of flavoured mesons produced in decays $\Phi^0 (1020)\rightarrow K^0 \bar K^0$, $\Psi(3770)\rightarrow D^0\bar D^0$, $\Upsilon(10580)\rightarrow B^0\bar B^0$ and $\Upsilon(10860)\rightarrow B^0_s\bar B^0_s$, provided the Lorentz boosts are large enough and experimental resolutions in momenta and time are sufficiently good.

\section*{Acknowledgment}
The work of WW is supported by the Polish National Science Centre project number 2017/26/M/ST2/00697.  For AB and  JKG, this work was supported by funds provided by the Polish National Science Center,  the project number  2019/33/B/ST2/00050.

\appendix
\section{Components of the boost matrix $L'_g$}\label{appA}

\begin{align}
		L_{00}
		=
		\frac{\gamma P_4 + P_0}{\gamma P_0 -\beta\gamma\mathbf{P}_1 + P_4}
		+
		(\gamma-1)
		\frac{\mathbf{P}^2}{P_+(\gamma P_0 -\beta\gamma\mathbf{P}_1 + P_4)}
		-
		\beta\gamma \frac{\mathbf{P}_1}{\gamma P_0 -\beta\gamma\mathbf{P}_1 + P_4}
\end{align}
\begin{align}
		L_{10}
		&=
		\frac{1}{\kappa}
		\Big[\gamma\mathbf{P}_1 - \beta\gamma P_4 - \beta\gamma\frac{\mathbf{P}^2}{P_+}
		-(\gamma\mathbf{P}_1 - \beta\gamma P_0)
		\Big(
		\frac{\gamma P_4 + P_0}{\gamma P_0 -\beta\gamma\mathbf{P}_1 + P_4}
		 \nonumber \\
        &+
		(\gamma-1)
		\frac{\mathbf{P}^2}{P_+(\gamma P_0 -\beta\gamma\mathbf{P}_1 + P_4)}
		-
		\beta\gamma \frac{\mathbf{P}_1}{\gamma P_0 -\beta\gamma\mathbf{P}_1 + P_4}
		\Big)
		\Big]
\end{align}
\begin{align}
		L_{20}
		&=
		\frac{\mathbf{P}_2}{\kappa}
		\Big[1-
		\Big(
		\frac{\gamma P_4 + P_0}{\gamma P_0 -\beta\gamma\mathbf{P}_1 + P_4}
		+
		(\gamma-1)
		\frac{\mathbf{P}^2}{P_+(\gamma P_0 -\beta\gamma\mathbf{P}_1 + P_4)}
        \nonumber \\
        &
		-
		\beta\gamma \frac{\mathbf{P}_1}{\gamma P_0 -\beta\gamma\mathbf{P}_1 + P_4}
		\Big)
		\Big]
\end{align}
\begin{align}
		L_{30}
		&=
		\frac{\mathbf{P}_3}{\kappa}
		\Big[1-
		\Big(
		\frac{\gamma P_4 + P_0}{\gamma P_0 -\beta\gamma\mathbf{P}_1 + P_4}
		+
		(\gamma-1)
		\frac{\mathbf{P}^2}{P_+(\gamma P_0 -\beta\gamma\mathbf{P}_1 + P_4)}
        \nonumber \\
        &
		-
		\beta\gamma \frac{\mathbf{P}_1}{\gamma P_0 -\beta\gamma\mathbf{P}_1 + P_4}
		\Big)
		\Big]
\end{align}
\begin{align}
		L_{01} 
		= 
		(\gamma-1) \kappa \frac{\mathbf{P}_1}{P_+(\gamma P_0 -\beta\gamma \mathbf{P}_1 + P_4)}
		-\frac{\beta\gamma\kappa}{\gamma P_0 -\beta\gamma \mathbf{P}_1 + P_4}
\end{align}
\begin{align}
		L_{11}
		=
		-\beta \gamma \frac{\mathbf{P}_1}{ P_+}
		+
		(1-\gamma) \frac{(\gamma\mathbf{P}_1 -\beta\gamma P_0) \mathbf{P}_1}{P_+(\gamma P_0 -\beta\gamma \mathbf{P}_1 + P_4)}
		+
		\frac{\beta\gamma (\gamma\mathbf{P}_1 - \beta\gamma P_0)}{\gamma P_0 -\beta\gamma \mathbf{P}_1 + P_4}
		+
		\gamma
\end{align}
\begin{align}
		L_{21} 
		=
		(1-\gamma) \frac{\mathbf{P}_1 \mathbf{P}_2}{P_+(\gamma P_0 -\beta\gamma \mathbf{P}_1 + P_4)}
		+
		\frac{\beta\gamma \mathbf{P}_2}{\gamma P_0 -\beta\gamma \mathbf{P}_1 + P_4}
\end{align}
\begin{align}
		L_{31} 
		=
		(1-\gamma) \frac{\mathbf{P}_1 \mathbf{P}_3}{P_+(\gamma P_0 -\beta\gamma \mathbf{P}_1 + P_4)}
		+
		\frac{\beta\gamma \mathbf{P}_3}{\gamma P_0 -\beta\gamma \mathbf{P}_1 + P_4}
\end{align}
\begin{align}
		L_{02} 
		= 
		(\gamma-1) \kappa \frac{\mathbf{P}_2}{P_+(\gamma P_0 -\beta\gamma \mathbf{P}_1 + P_4)}
\end{align}
\begin{align}
		L_{12}
		=
		-\beta \gamma \frac{\mathbf{P}_2}{ P_+}
		+
		(1-\gamma) \frac{(\gamma\mathbf{P}_1 -\beta\gamma P_0) \mathbf{P}_2}{P_+(\gamma P_0 -\beta\gamma \mathbf{P}_1 + P_4)}
\end{align}
\begin{align}
		L_{22} 
		=
		(1-\gamma) \frac{\mathbf{P}_2^2}{P_+(\gamma P_0 -\beta\gamma \mathbf{P}_1 + P_4)} + 1
\end{align}
\begin{align}
		L_{32} 
		=
		(1-\gamma) \frac{\mathbf{P}_2 \mathbf{P}_3}{P_+(\gamma P_0 -\beta\gamma \mathbf{P}_1 + P_4)}
\end{align}
\begin{align}
		L_{03} 
		= 
		(\gamma-1) \kappa \frac{\mathbf{P}_3}{P_+(\gamma P_0 -\beta\gamma \mathbf{P}_1 + P_4)}
\end{align}
\begin{align}
		L_{13}
		=
		-\beta \gamma \frac{\mathbf{P}_3}{ P_+}
		+
		(1-\gamma) \frac{(\gamma \mathbf{P}_1 -\beta\gamma P_0) \mathbf{P}_3}{P_+(\gamma P_0 -\beta\gamma \mathbf{P}_1 + P_4)}
\end{align}
\begin{align}
		L_{23} 
		=
		(1-\gamma) \frac{\mathbf{P}_2 \mathbf{P}_3}{P_+(\gamma P_0 -\beta\gamma \mathbf{P}_1 + P_4)}
\end{align}
\begin{align}
		L_{33} 
		=
		(1-\gamma) \frac{\mathbf{P}_3^2}{P_+(\gamma P_0 -\beta\gamma \mathbf{P}_1 + P_4)} + 1
\end{align}

\section{Explicit components of $L\triangleright P$ and $L'_g\triangleright Q$}\label{appB}

\begin{align}
		(L'_g\triangleright Q)_0
		&=
		\frac{1}{2} \gamma \Big(\beta \cos (\theta ) \sqrt{M^2-4 m^2} +\sqrt{5 M^2-4 m^2}\Big) \nonumber \\
        &+
        \frac{1}{40 \kappa }\Big[5 \beta \gamma \cos (\theta ) \sqrt{M^2-4 m^2} \Big(3 M-2 \sqrt{5 M^2-4 m^2}\Big) \nonumber \\ 
        &+
        4 m^2 \Big(5 \gamma ^2-\frac{12 \gamma M}{\sqrt{5 M^2-4 m^2}}-5\Big)+3 M \Big(\gamma \sqrt{5 M^2-4 m^2}-15 \Big(\gamma ^2-1\Big) M\Big) \nonumber \\
        &-
        5 (\gamma -1)^2 \cos (2 \theta ) \Big(4 m^2-M^2\Big)\Big]
        +O\Big(\Big(\frac{1}{\kappa }\Big)^2\Big)
		\\ \nonumber \\
		(L'_g\triangleright Q)_1
		&=
        -\frac{1}{2} \gamma \Big(\beta \sqrt{5 M^2-4 m^2}+\cos (\theta ) \sqrt{M^2-4 m^2}\Big) \nonumber \\
        &+
        \frac{1}{8 \kappa \sqrt{5 M^2-4 m^2}}\Big[\beta \gamma \Big((\gamma -2) \cos (2 \theta ) \Big(4 m^2-M^2\Big) \sqrt{5 M^2-4 m^2} \nonumber \\
        &+
        4 m^2 \Big(3 M-\gamma \sqrt{5 M^2-4 m^2}\Big)-3 M^2 \Big(M-3 \gamma \sqrt{5 M^2-4 m^2}\Big)\Big) \nonumber \\
        &+
        \cos (\theta ) \sqrt{M^2-4 m^2} \Big(M \Big(10 (\gamma -1) M-3 \gamma \sqrt{5 M^2-4 m^2}\Big) \nonumber \\
        &-8 (\gamma -1) m^2\Big)\Big]+O\Big(\Big(\frac{1}{\kappa }\Big)^2\Big)
		\\ \nonumber \\
		(L'_g\triangleright Q)_2
		&=
        \frac{1}{2} \sin (\theta ) \sqrt{M^2-4 m^2} \cos (\phi ) \nonumber \\
        &-\frac{1}{8 \kappa }\Big[\sin (\theta ) \sqrt{M^2-4 m^2} \cos (\phi ) \Big(2 \beta \gamma \cos (\theta ) \sqrt{M^2-4 m^2} \nonumber \\
        &+
        2 (\gamma -1) \sqrt{5 M^2-4 m^2}-3 M\Big)\Big]+O\Big(\Big(\frac{1}{\kappa }\Big)^2\Big)
		\\  \nonumber \\
		(L'_g\triangleright Q)_3
		&=
        \frac{1}{2} \sin (\theta ) \sqrt{M^2-4 m^2} \sin (\phi ) \nonumber \\
        &-
        \frac{1}{8 \kappa }\Big[\sin (\theta ) \sqrt{M^2-4 m^2} \sin (\phi ) \Big(2 \beta \gamma \cos (\theta ) \sqrt{M^2-4 m^2} \nonumber \\
        &+
        2 (\gamma -1) \sqrt{5 M^2-4 m^2}-3 M\Big)\Big]+O\Big(\Big(\frac{1}{\kappa }\Big)^2\Big)
	\end{align}

\begin{align}
    (L \triangleright P)_0
    &=
    \frac{1}{2} \gamma \left(\beta \cos (\theta ) \sqrt{M^2-4 m^2}+\sqrt{5 M^2-4 m^2}\right) \nonumber \\
    &+
    \frac{\gamma M \left(\beta \cos (\theta ) \sqrt{M^2-4 m^2} +\frac{M^2-4 m^2}{\sqrt{5 M^2-4 m^2}}\right)}{8 \kappa }+O\Big(\Big(\frac{1}{\kappa }\Big)^2\Big)
    \\ \nonumber \\
    (L \triangleright P)_1
    &=
    -\frac{1}{2} \gamma \left(\beta \sqrt{5 M^2-4 m^2}+\cos (\theta ) \sqrt{M^2-4 m^2}\right) \nonumber \\
    &+\frac{\gamma M \left(\frac{\beta \left(4 m^2-M^2\right)}{\sqrt{5 M^2-4 m^2}}-\cos (\theta ) \sqrt{M^2-4 m^2}\right)}{8 \kappa }+O\Big(\Big(\frac{1}{\kappa }\Big)^2\Big)
    \\ \nonumber \\
    (L \triangleright P)_2
    &=
    -\frac{1}{2} \sin (\theta ) \sqrt{M^2-4 m^2} \cos (\phi ) \nonumber \\
    &-\frac{M \sin (\theta ) \sqrt{M^2-4 m^2} \cos (\phi )}{8 \kappa }+O\Big(\Big(\frac{1}{\kappa }\Big)^2\Big)
    \\ \nonumber \\
    (L \triangleright P)_3
    &=
    -\frac{1}{2} \sin (\theta ) \sqrt{M^2-4 m^2} \sin (\phi ) \nonumber \\
    &-\frac{M \sin (\theta ) \sqrt{M^2-4 m^2} \sin (\phi )}{8 \kappa }+O\Big(\Big(\frac{1}{\kappa}\Big)^2\Big)
\end{align}
Notice that, in absence of deformation, the components of momenta along $x^2$ and $x^3$ (i.e. those components perpendicular to the boost direction) of $L'_g\triangleright Q$ and $L\triangleright P$ are equal and opposite. However, deformation introduces additional terms which make even these components not equal and opposite after boost, which in turn gives a deformed angular distribution.

\end{document}